# Anisotropic model with truncated linear dispersion for lattice and interfacial thermal transport in layered materials

**Hongkun Li, Weidong Zheng, Yee Kan Koh***

*Department of Mechanical Engineering, National University of Singapore, Singapore 117576*

*Corresponding Author. Email: mpekyk@nus.edu.sg

**ABSTRACT**

Recently, an anisotropic Debye model [Dames *et al.*, Physical Review B **87**, 12 (2013)] was proposed for calculations of the interfacial thermal conductance and the minimum thermal conductivity of graphite-like layered materials. Despite successes of the model in explaining heat transport mechanisms in layered materials (e.g., phonon focusing in highly anisotropic materials), the anisotropic Debye model assumes a phonon dispersion with unrealistic speeds of sounds especially for the flexural (ZA) phonons and overestimated cutoffs for all phonon branches. The deficiencies lead to substantially underestimated phonon irradiation for low-frequency phonons. Here, we develop an anisotropic model with truncated linear dispersion that resembles the real phonon dispersion, using speeds of sounds derived from elastic constants and cutoff frequencies derived from Brillouin zone boundaries. We also employ a piecewise linear function for the ZA phonons. Our model correctly calculates the phonon irradiation over a wide temperature range, verifying the accuracy of our model. We compare calculations of our and the Dames' models to measurements of thermal conductivity of graphite and thermal conductance of metal/graphite interfaces, and find that the two models differ significantly for heat transport across the basal planes in graphite even at high temperatures. Our work thus provides a convenient analytical tool to study the phonon transport properties in layered materials.

**TEXT**

In the past few years, van-der-Waals layered materials such as graphite, hexagonal BN, black phosphorus and transition-metal dichalcogenides (TMDs) attract widespread interest due to their unique electronic,[1,2] optical,[3,4] mechanical,[5] chemical[6] and thermal[7,8] properties. These unique properties lead to development of energy storage,[9] thermoelectrics,[10] radio-frequency transistors,[11] and optoelectronic[12,13] devices based on the layered materials. One unique property of the layered materials is poor interfacial and lattice thermal transport across the basal planes due to the weak van-der-Waals bondings.[8,14] The unique thermal property is an important consideration in the designs of devices of layered materials. On one hand, the low thermal conductance across the basal planes could limit the performance of the nanodevices of layered materials. For example, heat dissipation from hot spots in graphene-based nanodevices is limited by the poor thermal conductance of graphene interfaces,[15-18] despite the high thermal conductivity along the basal planes. On the other hand, the poor thermal properties could lead to better performance of thermoelectric materials, in which low thermal conductivity enhances the thermoelectric figure of merit (ZT) and thus the efficiency of thermoelectric energy conversion. The excellent ZT value recently reported in SnSe crystal has been attributed to its low thermal conductivity.[19]

Despite the importance, accurate and simple models for phonon transport in layered materials are still lacking. Historically, thermal transport properties (e.g., the thermal conductivity and interfacial thermal conductance) are approximated using Debye-Callaway type continuum models,[20-22] in which phonon relaxation times and transmission probability are treated empirically as fitting parameters. (In recent years, the phonon relaxation times have been calculated from first-principles calculations.[23,24]) In the Debye-Callaway type models, how to approximate the dispersion relation seriously affects the accuracy of the models. For materials with isotropic thermal conductivity (e.g., Si and a lot of common metals), the thermal conductivity and the thermal conductance have been approximated using various dispersion relations, e.g., lattice dynamics calculations,[25] sine-type approximation (Born–von Karman),[26] truncated linear dispersion[27,28] and dispersion fitted from measurements[29]. However, for graphite-like layered materials, phonon dispersion varies significantly from the *ab*-plane (the basal plane) to the *c*-axis. As a result, two key assumptions in the isotropic Debye-like models, i.e., spherical isoenergy energy surface and spherical first Brillouin zone (FBZ), are no longer valid. In addition, a characteristic feature of the dispersion relation of layered materials is that the dispersion of the ZA branch (the flexural phonon branch) is often complex, exhibiting low group velocities for phonons at the center ($\Gamma$ point) but high group velocities for phonons near the edge of the FBZ. Thus, how to approximate these unique features of the phonon dispersion of layered materials in a simple yet physically meaningful model is a serious challenge for researchers.

Recently, Dames *et al.*[30] developed an anisotropic Debye model for computation of the interfacial thermal conductance (based on the phonon radiation limit[26] and the diffuse mismatch model[22,31]) and the minimum anisotropic thermal conductivity (e.g., $MoS_2$[7], $WSe_2$[32]) of layered materials. To model the complex phonon dispersion, the anisotropic Debye model by Dames *et al.* assumes that the FBZ boundary and the isoenergy surface are ellipsoids. With the anisotropic models, Dames *et al.*[32] successfully explained the origins of ultralow thermal conductivity across the basal planes observed in disorder, layered materials. Despite the successes, in Dames' model, oversimplified "secant" sound velocities (i.e., the slope of a secant line connecting $\Gamma$ point and the FBZ boundary of the same branch) and overestimated cutoffs were employed for all three polarizations of acoustic phonons (i.e., longitudinal (LA), transverse (TA) and flexural (ZA) acoustic phonons), see Fig. 1(a).

The simple "secant" sound velocities are not a good representation of the phonon dispersion of ZA branch, and thus fail to capture the real group velocities of low-frequency phonons. Since low-frequency acoustics phonons carry a significant amount of heat due to their long mean free paths,[33,34] the unrealistic sound velocities lead to a large discrepancy for phonon irradiation between the anisotropic Debye model and lattice dynamics calculations at low temperatures, see Fig. 12 in Ref. 30.

In this paper, we develop a new anisotropic model with a realistic phonon dispersion that captures the important features of phonon dispersions of layered materials. For the LA and TA branches, we assume a truncated linear dispersion that gives the measured speeds of sounds in the low-frequency limit and is truncated at cutoff frequencies derived from phonon frequencies at the first Brillouin zone boundaries. For the ZA branch, we apply a piecewise linear function to approximate the unique phonon dispersion. Thus, the proposed model is capable of capturing the phonon irradiation at both low and high temperatures accurately. Our model is simple to implement and could be used by experimentalists to quickly evaluate their experimental results.

Lattice and interfacial thermal transport by phonons depend on both the properties of crystal structures (i.e., the phonon velocity and density of states) and phonon transport properties (e.g., the relaxation times or the transmission probability). For the phonon transport in crystals, the thermal conductivity in any $s$-direction ($\Lambda_s$) can be expressed as in Eq. (1).[32] (In this paper, we define $\Lambda_s = -q_s / \Delta T_s$, where $q_s$ and $\Delta T_s$ are heat flux and temperature difference in $s$-direction, respectively. Thus, when $s$ refers to the directions along $ab$-plane and $c$-axis, $\Lambda_s$ denotes the through-plane and in-plane components of the thermal conductivity tensor, respectively.)

$$\Lambda_s = \sum_j \int_\omega \hbar \omega g_{s,j}(\omega) \frac{\partial f_{BE}}{\partial T} \tau_j(\omega) d\omega \qquad (1)$$

$$g_{s,j}(\omega) = \frac{1}{8\pi^3} \iint \frac{(\mathbf{v}_{g,j} \cdot \hat{\mathbf{s}})^2}{\|\mathbf{v}_{g,j}\|} dS_\omega \qquad (2)$$

where subscript $j$ denotes the three phonon polarizations of acoustic phonons, $g_{s,j}(\omega)$ is sum of squares of $\mathbf{v}_{g,j}$ components in the $s$-direction for all phonons of frequency $\omega$ in branch $j$, $\mathbf{v}_{g,j}$ is the group velocity, $\tau_j(\omega)$ is the relaxation time that mostly depends on $\omega$ and is usually fitted empirically or calculated from the first principles models, $dS_\omega$ is an elemental area on an isoenergy surface of frequency $\omega$ in the FBZ in the reciprocal space, $f_{BE}$ is the Bose-Einstein distribution function, and $\hbar$ is the reduced Planck's constant. (Here, we assume that all optical phonons are non-propagating with zero group velocity, and thus do not contribute to heat transport. The polarization $j$ usually refers to LA, TA or ZA branches, but in our final implementation, we use $j$ to refer to TA, TL1 and TL2 branches, see the discussion below.)

On the other hand, for the interface of materials 1 and 2 in which $s$-direction is perpendicular to the interface, the interfacial thermal conductance ($G$) can be expressed as[22]

$$G = \sum_j \int_\omega \hbar\omega \left[ h_{s,j}(\omega) \right]_1 \frac{\partial f_{BE}}{\partial T} \alpha_{1-2}(\omega) d\omega \qquad (3)$$

$$h_{s,j}(\omega) = \frac{1}{8\pi^3} \iint_{\mathbf{k}\cdot\hat{\mathbf{s}}<0} \frac{(\mathbf{v}_{g,j} \cdot \hat{\mathbf{s}})}{\|\mathbf{v}_{g,j}\|} dS_\omega \qquad (4)$$

where $h_{s,j}(\omega)$ is the phonon flux of phonons of frequency $\omega$ and branch $j$ in the $s$-direction, the subscript "1" here refers to material 1, $\alpha_{1-2}(\omega)$ is transmission probability from material 1 to material 2, $\mathbf{k} = (k_a, k_b, k_c)$ is the wavevector (see Fig. 3 for the directions of $k_a$, $k_b$ and $k_c$ axes), $k_{ab}^2 = k_a^2 + k_b^2$, and $\mathbf{k} \cdot \hat{\mathbf{s}} < 0$ denotes integration over half of the incident FBZ in material 1.

In Eqs. (1) and (3), we group the properties of crystal structures (i.e., phonon group velocity and density of states, DOS) into two terms, $g_{s,j}(\omega)$ and $h_{s,j}(\omega)$, which are expressed in Eqs. (2) and

(4) in a general form that can be applied to any phonon dispersion. We note that $g_{s,j}(\omega)$ and $h_{s,j}(\omega)$ are also known as v²DOS and vDOS, respectively, in some other literature.[30,32] The role of $g_{s,j}(\omega)$ and $h_{s,j}(\omega)$ in Eqs. (1) and (2) is analogous to the role of the DOS in the specific heat. Moreover, $h_{s,j}(\omega)$ and the total phonon flux of frequency ω, $h_s(\omega) = \sum_j h_{s,j}(\omega)$, are often involved in the evaluation of the transmission probability $\alpha_{1-2}(\omega)$. For examples, in the diffuse mismatch model (DMM),[22] $\alpha_{1-2}(\omega) = [h_s(\omega)]_2/([h_s(\omega)]_1 + [h_s(\omega)]_2)$, see the explanation of the expression below, while in the phonon radiation limit,[26] $\alpha_{1-2}(\omega) = [h_s(\omega)]_2/[h_s(\omega)]_1$ if $[h_s(\omega)]_2 < [h_s(\omega)]_1$, see the discussion on this expression in Ref. 35. Thus, accurate calculations of $g_{s,j}(\omega)$ and $h_{s,j}(\omega)$ are crucial for calculations of the anisotropic thermal conductivity and interfacial thermal conductance.

In the following text, we focus on evaluating Eqs. (2) and (4) for $g_{s,j}(\omega)$ and $h_{s,j}(\omega)$ of layered materials. To achieve this goal, we first approximate the highly anisotropic phonon dispersion for layered materials using a simple and realistic dispersion, with two independent parameters ($v_{j,i}$ and $\omega_{j,i}$) for each of the LA and TA branches in either *ab*-plane or *c*-axis, and four independent parameters ($v_{Z,1}$, $v_{Z,2}$, $\omega_{Z,1}$, $\omega_{Z,2}$) for the ZA branch. (Here, the first subscript *j* denotes either LA (L), TA (T) or ZA (Z) branches, and the second subscript *i* denotes either the direction for the LA and TA branches or segment number ("1" or "2") for the ZA branch, see below.) For LA and TA phonons, we assume a truncated linear dispersion, in which phonons of wavevector $k < k_{j,i}$ of the polarization *j* (LA or TA) along the direction *i* (*ab*-plane or *c*-axis) are assumed to have a constant speed of sounds of $v_{j,i}$, while phonons of $k > k_{j,i}$ are assumed to be non-propagating, see Fig. 1(a). We derive $v_{j,i}$ from the elastic constants, see Table 3 for the equations for the speeds of sounds that we employ. We then set the cutoff wavevectors ($k_{j,i}$) from the cutoff frequencies ($\omega_{j,i}$) using $\omega_{j,i} = v_{j,i}k_{j,i}$. The cutoff frequencies $\omega_{j,ab}$ along *ab*-plane is obtained from the average of phonon frequencies at the high-symmetry M and K points at the FBZ boundaries, and $\omega_{j,c}$ along *c*-axis from the phonon frequency at A' point, see Fig. 1(a). Note that here we follow Ref. 30 to unfold the dispersion relation along the *c*-axis direction, considering the relatively high velocity of optical phonons along the *c*-direction.

For ZA phonons, we employ a piecewise linear function, in which we divide the dispersion into two linear segments, up to two cutoff points of $(k_{Z,1}, \omega_{Z,1})$ and $(k_{Z,2}, \omega_{Z,2})$, see Fig. 2. Phonons of $k < k_{Z,1}$ are approximated to have a speed of sound $v_{Z,1}$, while phonons of $k > k_{Z,1}$ are assumed to have a fitted group velocity of $v_{Z,2}$. To set the four independent parameters for the ZA branch, we first derive $v_{Z,1}$ from the speed of sound of ZA phonons, see Table 3 for the corresponding equation. Then, we derive the second cutoff point $(k_{Z,2}, \omega_{Z,2})$ from the averages of the phonon frequencies and wavevectors at points M and K at the FBZ boundaries, see Table 3. We adjust the $v_{Z,2}$ such that our piecewise linear dispersion matches the experimental dispersion of the layered material, see Fig. 1(a). The first cutoff point $(k_{Z,1}, \omega_{Z,1})$ can then be determined from the intersection of two lines in our piecewise linear dispersion.

We compare the dispersion assumed in this work and in the anisotropic model by Dames *et al.*[30] in Fig. 1(a). We find that for long-wavelength phonons, the phonon dispersion we assume using the speeds of sounds matches the measured dispersion well, while the phonon dispersion with the "secant" group velocities assumed in the anisotropic model of Dames *et al.* deviates considerably. For example, for graphite, the value of $v_{Z,1}$ in our model approaches the speed of sound of ZA phonons while the "secant" velocity of ZA phonons assumed in the Dames' model is 4× larger, as summarized in Table 3. The overestimation of the group velocity of ZA phonons leads to underestimation of the heat transport in graphite by the Dames' model, see our discussion below. On the other hand, for short-wavelength phonons, we exclude the phonon modes near the FBZ boundaries to heat transport using zero group velocity, see Fig. 1(a), while Dames *et al.* assume all phonons (including short-wavelength phonons near FBZ boundaries) contribute substantially to heat transport with the same "secant" group velocity. Thus, the Dames' model overestimates the contribution of short-wavelength phonons. We note that with the truncated linear dispersion, we do not use the total number of acoustic phonons in LA and TA branches to set the cutoff frequencies $\omega_{j,i}$, and thus only include a portion of acoustic phonons in the calculations of thermal transport. (As a result, the truncated linear dispersion is not suitable for calculations of heat capacity, because with the assumption, the Dulong-Petit limit is not approached at high temperatures.) We think that the truncated linear approximation is a more realistic approximation for calculations of transport properties because short-wavelength phonons near the Brillouin zone boundaries, while contribute dominantly to lattice heat capacity, do not substantially contribute to heat transport.

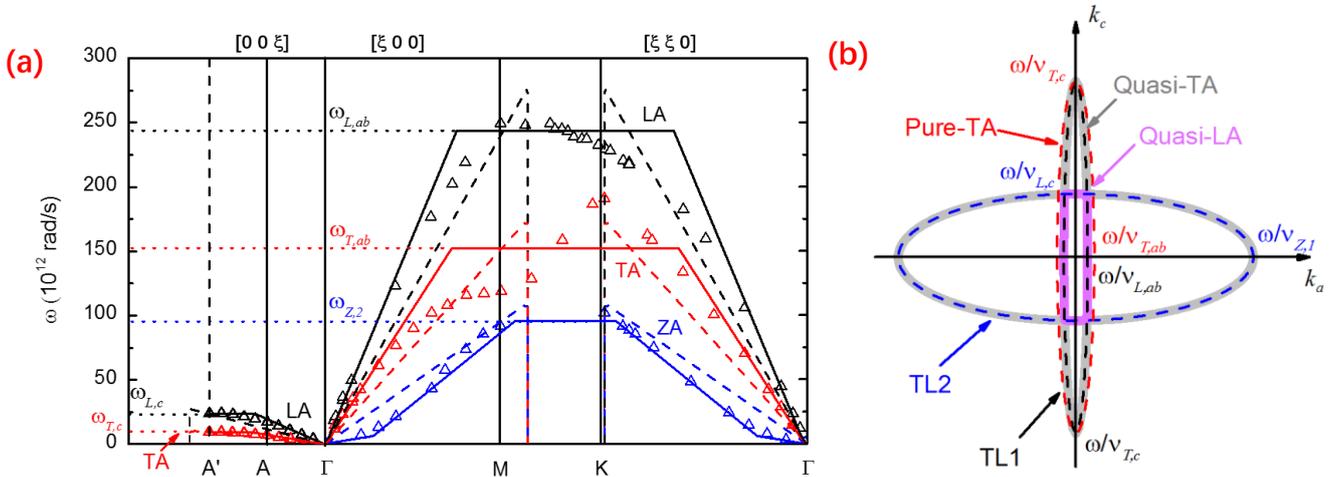

**Figure 1. a)** Truncated linear dispersion (solid lines) of graphite assumed in this work, compared to the dispersion assumed by Ref. 30 (dashed lines), for all three branches in phonon dispersion of graphite[36] along the *c*-axis (Γ-A direction) and the *ab*-plane (Γ-M and Γ-K directions). A' is the high symmetric point after unfolding the phonon dispersion along the *c*-axis. The open symbols are experimental measurements of the phonon dispersion of graphite.[37] **b)** Cross-section of the isoenergy surfaces of frequency ω on $k_a$-$k_b$ plane, for quasi-TA (lobed, light gray), quasi-LA (nearly rectangular, violet) and pure-TA (ellipse, red dashed line) branches, calculated according to a continuum elasticity theory.[38] In our model, we recompose the quasi-LA and quasi-TA branches into TL1 (black dashed line) and TL2 (blue dashed line) branches.

In the evaluation of Eqs. (1) and (3), we do not sum the heat transport over conventional branches of LA, TA and ZA phonons. Instead, we follow Dames *et al.*[30] to recompose the quasi-LA and quasi-TA branches into TL1 and TL2 branches. The reason for the recomposition can be better understood by observing the angle dependent dispersion ($k/\omega$) of highly anisotropic materials, see the Fig. 1(b) for the cross section of isoenergy surfaces of long-wavelength phonons in graphite in the $k_a$-$k_c$-plane calculated from a continuum elasticity theory.[38] We note the different shapes of the isoenergy surfaces for the three branches, the lobed quasi-TA (light gray), the rectangular quasi-LA (violet) and the ellipsoidal pure-TA (red dashed) branches, see Fig. 1(b). In the evaluation of the $g_{s,j}(\omega)$ and $h_{s,j}(\omega)$ in Eqs. (2) and (4), we find that while integration over the ellipsoidal isoenergy surfaces of the pure-TA branch can be readily performed, integration over the lobed isoenergy surface of the quasi-TA branch is challenging. To simplify the integration over the isoenergy surfaces, we thus follow Dames *et al.*[30] to decompose the quasi-TA and quasi-LA branches in Fig. 1(b) and recompose the isoenergy surfaces into two ellipsoidal TL1 (black dashed line) and TL2 (blue dashed line) branches. The TL1 branch has major and minor axes of $\omega/v_{T,c}$ (TA phonons along *c*-axis) and $\omega/v_{L,ab}$ (LA phonons in *ab*-plane), respectively, while the TL2 branch has major and minor axes of $\omega/v_{Z,1}$ (ZA phonons in *ab*-plane) and $\omega/v_{L,c}$ (LA phonons along *c*-axis), respectively.

Our approach is different from the anisotropic Debye model by Dames *et al.*[30] particularly for our treatment of the TL2 branch, since we assume a two-segment, piecewise linear dispersion for the ZA phonons. To evaluate $g_{s,j}(\omega)$ and $h_{s,j}(\omega)$ for the TL2 branch, we first consider the isoenergy surface and the effective FBZ boundaries of the TL2 branch. For phonons of frequency $\omega$, taking into consideration the piecewise linear dispersion of the ZA branch and assuming that the dispersion is isotropic in *ab*-plane, the isoenergy surface can be expressed as

$$\begin{cases} \dfrac{v_{Z,2}^2 k_{ab}^2}{\omega^2} + \dfrac{v_{L,c}^2 k_c^2}{\omega^2} = 1, & \omega \leq \omega_{Z,1} \\ \dfrac{v_{Z,2}^2 k_{ab}^2}{(\omega - \Delta\omega)^2} + \dfrac{v_{L,c}^2 k_c^2}{\omega^2} = 1, & \omega_{Z,1} < \omega < \omega_{Z,2} \end{cases} \quad (5)$$

where $\Delta\omega = k_{Z,1}(v_{Z,1} - v_{Z,2})$. The derivation of Eq. (5) is presented in the Supplementary.

We then define an effective FBZ by setting that phonons within the effective FBZ have non-zero group velocity, while phonons outside of the FBZ are non-propagating. For the truncated linear dispersion that we assume, the effective FBZ for all branches are ellipsoids, with the boundaries defined by $k_{j,i}$ for LA and TA phonons and $k_{Z,2}$ for ZA phonons. Mathematically, the effective FBZ boundaries of the TL2 branch are defined by[32,39]

$$\frac{k_{ab}^2}{k_{Z,2}^2} + \frac{k_c^2}{k_{L,c}^2} = 1 \quad (6)$$

We evaluate the surface integral of Eqs. (2) and (4) in *ab*-plane and *c*-axis direction over isoenergy surface $S_\omega$ of the TL2 branch by projecting the isoenergy surface to $k_a - k_b$ plane. We apply the polar coordinate substitution ($k_a = \rho \cos\varphi$, $k_b = \rho \sin\varphi$) to simplify the evaluation. The domain of polar angle $\varphi$ can be easily set: for $h_{c,j}(\omega)$ and all $g_{s,j}(\omega)$, we apply $0 \leq \varphi \leq 2\pi$; for $h_{s,j}(\omega)$ where *s* is in the *ab*-plane, we use $-\pi/2 \leq \varphi \leq \pi/2$. The domain of the polar radius $\rho$, however, depends on the value of ω, see Fig. 2. If $\omega < \omega_{Z,1}$, the isoenergy surface lies fully within the effective FBZ, we thus set $0 \leq \rho \leq \omega/v_{Z,1}$. If $\omega_{Z,1} < \omega < \omega_{L,c}$, the isoenergy surface is still fully within the effective FBZ, but the limit of the integration is slightly different as ZA phonons are now within the second segment of the piecewise dispersion; we thus set $0 \leq \rho \leq (\omega - \Delta\omega)/v_{Z,2}$. If $\omega_{L,c} < \omega < \omega_{Z,2}$, parts of the isoenergy surface lie outside of the effective FBZ, and thus should be excluded from the integration. By carefully consider the projection of isoenergy surface inside of the effective FBZ on $k_a - k_b$ plane, we then set $\rho_{min} \leq \rho \leq \rho_{max}$, where

$$\rho_{min} = k_{Z,2}\sqrt{\frac{\omega^2(\omega-\Delta\omega)^2 - \omega_{L,c}^2(\omega-\Delta\omega)^2}{k_{Z,2}^2 v_{Z,2}^2 \omega^2 - \omega_{L,c}^2(\omega-\Delta\omega)^2}} \tag{7}$$

$$\rho_{max} = (\omega - \Delta\omega)/v_{Z,2} \tag{8}$$

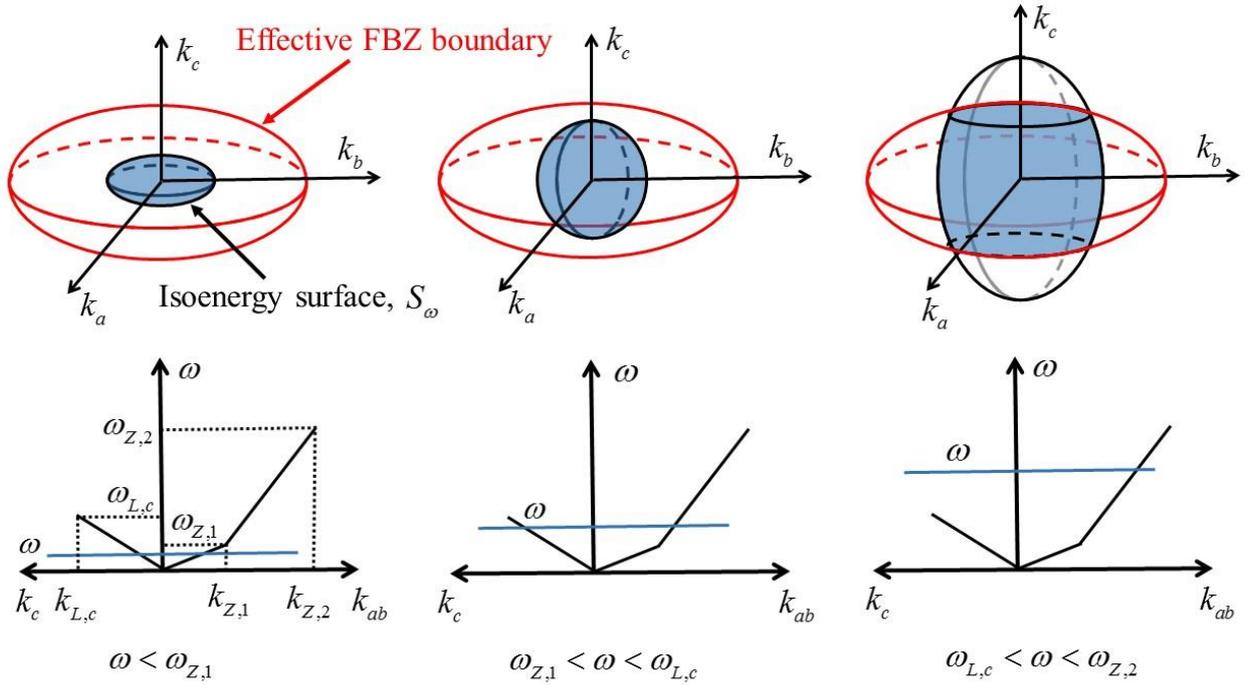

**Figure 2.** Three frequency regimes for TL2 branch. The red ellipsoid is the effective FBZ boundary, while the black one is the isoenergy surface for phonon frequency of ω. The blue shading on the isoenergy surface indicates the region within the effective FBZ, in which phonons are considered in the evaluation of Eqs. (2) and (4).

With the identified domains of $\varphi$ and $\rho$, we solve Eqs. (2) and (4) for $g_{s,j}(\omega)$ and $h_{s,j}(\omega)$ for the TL2 branch, for $s$ in the *ab*-plane and along *c*-axis, see the Supplementary Material for the details of the derivation. We summarize the derived equations in Table 1 and Table 2. The equations can be easily extended for the TA and TL1 branches, by setting $v_{Z,1} = v_{Z,2}$. We thus also list the equations for the TA and TL1 branches in Table 1 and Table 2.

Table 1. Expressions of $g_{s,j}(\omega)$ in *ab*-plane and along *c*-axis for all branches (TA, TL1 and TL2)

| Branches | Frequency regimes | $g_{s,j}(\omega)$ | |
|---|---|---|---|
| | | *c*-axis | *ab*-plane |
| TA | $\omega < \omega_{T,c}$ | $\dfrac{v_{T,c}}{6\pi^2 v_{T,ab}^2}\omega^2$ | $\dfrac{1}{6\pi^2 v_{T,c}}\omega^2$ |
| | $\omega_{T,c} < \omega < \omega_{T,ab}$ | $\dfrac{v_{T,c}}{6\pi^2 v_{T,ab}^2}\dfrac{1}{\omega}\left[\dfrac{\omega_{T,c}^2(\omega_{T,ab}^2-\omega^2)}{\omega_{T,ab}^2-\omega_{T,c}^2}\right]^{\frac{3}{2}}$ | $\dfrac{\omega_{T,c}\omega}{4\pi^2 v_{T,c}}\left[\dfrac{\omega_{T,ab}^2-\omega^2}{\omega_{T,ab}^2-\omega_{T,c}^2}\right]^{\frac{1}{2}} - \dfrac{\omega_{T,c}^3}{12\pi^2 v_{T,c}\omega}\left[\dfrac{\omega_{T,ab}^2-\omega^2}{\omega_{T,ab}^2-\omega_{T,c}^2}\right]$ |
| TL1 | $\omega < \omega_{T,c}$ | $\dfrac{v_{T,c}}{6\pi^2 v_{L,ab}^2}\omega^2$ | $\dfrac{1}{6\pi^2 v_{T,c}}\omega^2$ |
| | $\omega_{T,c} < \omega < \omega_{L,ab}$ | $\dfrac{v_{T,c}}{6\pi^2 v_{L,ab}^2}\dfrac{1}{\omega}\left[\dfrac{\omega_{T,c}^2(\omega_{L,ab}^2-\omega^2)}{\omega_{L,ab}^2-\omega_{T,c}^2}\right]^{\frac{3}{2}}$ | $\dfrac{\omega_{T,c}\omega}{4\pi^2 v_{T,c}}\left[\dfrac{\omega_{L,ab}^2-\omega^2}{\omega_{L,ab}^2-\omega_{T,c}^2}\right]^{\frac{1}{2}} - \dfrac{\omega_{T,c}^3}{12\pi^2 v_{T,c}\omega}\left[\dfrac{\omega_{L,ab}^2-\omega^2}{\omega_{L,ab}^2-\omega_{T,c}^2}\right]$ |
| TL2 | $\omega < \omega_{Z,1}$ | $\dfrac{v_{L,c}}{6\pi^2 v_{Z,1}^2}\omega^2$ | $\dfrac{1}{6\pi^2 v_{L,c}}\omega^2$ |
| | $\omega_{Z,1} < \omega < \omega_{L,c}$ | $\dfrac{v_{L,c}(\omega-\Delta\omega)^2}{2\pi^2}\int \dfrac{\rho\sqrt{(\omega-\Delta\omega)^2-\rho^2 v_{Z,2}^2}}{\rho^2 v_{Z,2}^2 \Delta\omega+(\omega-\Delta\omega)^3}d\rho$ | $\dfrac{1}{4\pi^2}\int \dfrac{\rho^3 v_{Z,2}^4 \omega^2}{\rho^2 v_{Z,2}^2\Delta\omega+(\omega-\Delta\omega)^3}\dfrac{1}{v_{L,c}\sqrt{(\omega-\Delta\omega)^2-\rho^2 v_{Z,2}^2}}d\rho$ |
| | $\omega_{L,c} < \omega < \omega_{Z,2}$ | $\dfrac{v_{L,c}(\omega-\Delta\omega)^2}{2\pi^2}\int \dfrac{\rho\sqrt{(\omega-\Delta\omega)^2-\rho^2 v_{Z,2}^2}}{\rho^2 v_{Z,2}^2 \Delta\omega+(\omega-\Delta\omega)^3}d\rho$ | $\dfrac{1}{4\pi^2}\int \dfrac{\rho^3 v_{Z,2}^4 \omega^2}{\rho^2 v_{Z,2}^2\Delta\omega+(\omega-\Delta\omega)^3}\dfrac{1}{v_{L,c}\sqrt{(\omega-\Delta\omega)^2-\rho^2 v_{Z,2}^2}}d\rho$ |

Table 2. Expressions of $h_{s,j}(\omega)$ in *ab*-plane and along *c*-axis for all branches (TA, TL1 and TL2)

| Branches | Frequency regimes | $h_{s,j}(\omega)$ | |
|---|---|---|---|
| | | *c*-axis | *ab*-plane |
| TA | $\omega < \omega_{T,c}$ | $\dfrac{1}{8\pi^2 v_{T,ab}^2}\omega^2$ | $\dfrac{1}{8\pi^2 v_{T,ab} v_{T,c}}\omega^2$ |
| | $\omega_{T,c} < \omega < \omega_{T,ab}$ | $\dfrac{1}{8\pi^2 v_{T,ab}^2}\left[\omega_{T,c}^2\left(\dfrac{\omega_{T,ab}^2-\omega^2}{\omega_{T,ab}^2-\omega_{T,c}^2}\right)\right]$ | $\dfrac{1}{2\pi^3}\dfrac{v_{T,ab}^2}{v_{T,c}}\int \dfrac{\rho^2}{\sqrt{(\omega^2-\rho^2 v_{T,ab}^2)}}d\rho$ |
| TL1 | $\omega < \omega_{T,c}$ | $\dfrac{1}{8\pi^2 v_{L,ab}^2}\omega^2$ | $\dfrac{1}{8\pi^2 v_{L,ab} v_{T,c}}\omega^2$ |
| | $\omega_{T,c} < \omega < \omega_{L,ab}$ | $\dfrac{1}{8\pi^2 v_{L,ab}^2}\left[\omega_{T,c}^2\left(\dfrac{\omega_{L,ab}^2-\omega^2}{\omega_{L,ab}^2-\omega_{T,c}^2}\right)\right]$ | $\dfrac{1}{2\pi^3}\dfrac{v_{L,ab}^2}{v_{T,c}}\int \dfrac{\rho^2}{\sqrt{(\omega^2-\rho^2 v_{L,ab}^2)}}d\rho$ |
| TL2 | $\omega < \omega_{Z,1}$ | $\dfrac{1}{8\pi^2 v_{Z,1}^2}\omega^2$ | $\dfrac{1}{8\pi^2 v_{Z,1} v_{L,c}}\omega^2$ |
| | $\omega_{Z,1} < \omega < \omega_{L,c}$ | $\dfrac{1}{8\pi^2 v_{Z,2}^2}(\omega-\Delta\omega)^2$ | $\dfrac{1}{8\pi^2 v_{Z,2} v_{L,c}}\omega(\omega-\Delta\omega)$ |
| | $\omega_{L,c} < \omega < \omega_{Z,2}$ | $\dfrac{1}{8\pi^2}\dfrac{(\omega-\Delta\omega)^2}{v_{Z,2}^2}\dfrac{v_{Z,2}^2 k_{Z,c}^2 \omega_{L,c}^2-(\omega-\Delta\omega)^2\omega_{L,c}^2}{k_{Z,2}^2 v_{Z,2}^2 \omega^2-\omega_{L,c}^2(\omega-\Delta\omega)^2}$ | $\dfrac{1}{2\pi^3}\dfrac{v_{Z,2}^2}{v_{L,c}}\int \dfrac{\rho^2\omega}{\sqrt{(\omega-\Delta\omega)^2-\rho^2 v_{Z,2}^2}(\omega-\Delta\omega)}d\rho$ |

We verify the accuracy of our approach by comparing the total phonon irradiation of graphite along the *c*-axis to calculations by a lattice dynamics model presented in Ref. 40. The phonon irradiation along the *c*-axis can be expressed as[32]

$$H_c = \sum_j \int_\omega \hbar\omega h_{c,j}(\omega) f_{BE} d\omega \qquad (9)$$

We derive the properties for graphite from the phonon dispersion, see Table 3. We compare our calculations of phonon irradiation of graphite along the c-axis, to calculations of a lattice dynamics model and the anisotropic Debye model by Dames *et al.*[30] in Fig. 3. We find that calculations of our model agree very well with the lattice dynamic calculations, over the whole temperature range. Note that all parameters in our model are derived from the phonon dispersion. On the other hand, calculations of the anisotropic Debye model by Dames *et al.* deviate significantly from the lattice dynamic calculations at low temperatures. The disagreement between calculations of model by Dames *et al.* and calculations of lattice dynamic originates from the overestimated group velocity for ZA phonons at low frequency and "secant" velocities used for other two branches. The comparison in Fig. 3 suggests that our approach is capable of predicting the phonon flux over the whole frequency range.

Table 3. Input parameters for graphite and $MoS_2$ (plots can be found in Supplementary Material) determined from published phonon dispersion. $C_{11}$, $C_{33}$, $C_{44}$ and $C_{66}$ are elastic constants, and $[\omega]_{j,M}$, $[\omega]_{j,K}$ and $[\omega]_{j,A'}$ refer to the frequency of phonons of branch $j$ at high-symmetry points M, K, and A' respectively, $[k]_{j,M}$ and $[k]_{j,K}$ refer to the wavevector of branch $j$ at points M and K respectively. For Dames' model, symbols are same with that in Ref. [30], $k_{c,max}$ and $k_{ab,max}$ are the maximum wavevectors along $c$-axis and $ab$-plane that are determined by ensuring correct acoustic modes, referring to Eqs. (3) and (20) in Ref. 30. Velocities of graphite in Dames' model are extracted directly from Table 3 in Ref. [30]. Velocities of $MoS_2$ are derived by "secant" method along Γ-K direction.

| Branch | This work | | | | Anisotropic model by Dames et al.[30] | | |
|---|---|---|---|---|---|---|---|
| | Properties | Equations | Graphite[37,41] | $MoS_2$[7,42] | Properties[30] | Graphite[30] | $MoS_2$ |
| TA | $v_{T,c}$ (m/s) | $v_{T,c}=(C_{44}/\rho)^{0.5}$ | 1487 | 1938 | $v_{c,TA}$ (m/s) | 1000 | 1519 |
| | $v_{T,ab}$ (m/s) | $v_{T,ab}=(C_{66}/\rho)^{0.5}$ | 14236 | 5372 | $v_{ab,TA}$ (m/s) | 10200 | 2209 |
| | $\omega_{T,c}$ ($10^{12}$ rad/s) | $[\omega]_{T,A'}$ | 8.14 | 7.77 | $k_{c,max}v_{c,TA}$ ($10^{12}$ rad/s) | 11.0 | 7.32 |
| | $\omega_{T,ab}$ ($10^{12}$ rad/s) | $([\omega]_{T,M}+[\omega]_{T,K})/2$ | 162 | 30.7 | $k_{ab,max}v_{ab,TA}$ ($10^{12}$ rad/s) | 176.5 | 24.0 |
| TL1 | $v_{T,c}$ (m/s) | $v_{T,c}=(C_{44}/\rho)^{0.5}$ | 1487 | 1938 | $v_{c,TL1}$ (m/s) | 1000 | 1519 |
| | $v_{L,ab}$ (m/s) | $v_{L,ab}=(C_{11}/\rho)^{0.5}$ | 22152 | 6850 | $v_{ab,TL1}$ (m/s) | 16200 | 3342 |
| | $\omega_{T,c}$ ($10^{12}$ rad/s) | $[\omega]_{T,A'}$ | 8.14 | 7.77 | $k_{c,max}v_{c,TL1}$ ($10^{12}$ rad/s) | 11.0 | 7.32 |
| | $\omega_{L,ab}$ ($10^{12}$ rad/s) | $([\omega]_{L,M}+[\omega]_{L,K})/2$ | 252 | 44.5 | $k_{ab,max}v_{ab,TL1}$ ($10^{12}$ rad/s) | 280.26 | 36.2 |
| TL2 | $v_{L,c}$ (m/s) | $v_{L,c}=(C_{33}/\rho)^{0.5}$ | 4138 | 3206 | $v_{c,TL2}$ (m/s) | 2500 | 2533 |
| | $v_{Z,1}$ (m/s) | $v_{Z,1}=(C_{44}/\rho)^{0.5}$ | 1487 | 1938 | $v_{ab,TL2}$ (m/s) | 6400 | 2563 |
| | $v_{Z,2}$ (m/s) | fitted | 7170 | 3010 | | | |
| | $\omega_{L,c}$ ($10^{12}$ rad/s) | $[\omega]_{L,A'}$ | 22.4 | 12.8 | $k_{c,max}v_{c,TL2}$ ($10^{12}$ rad/s) | 27.5 | 12.2 |
| | $\omega_{Z,1}$ ($10^{12}$ rad/s) | fitted | 5.20 | 5.54 | $k_{ab,max}v_{ab,TL2}$ ($10^{12}$ rad/s) | 110.7 | 27.7 |
| | $\omega_{Z,2}$ ($10^{12}$ rad/s) | $([\omega]_{Z,M}+[\omega]_{Z,K})/2$ | 94.9 | 34.1 | | | |
| | $k_{Z,2}$ ($10^{10}$ m$^{-1}$) | $([k]_{Z,M}+[k]_{Z,K})/2$ | 1.60 | 1.25 | $k_{ab,max}$ ($10^{10}$ m$^{-1}$) | 1.73 | 1.08 |

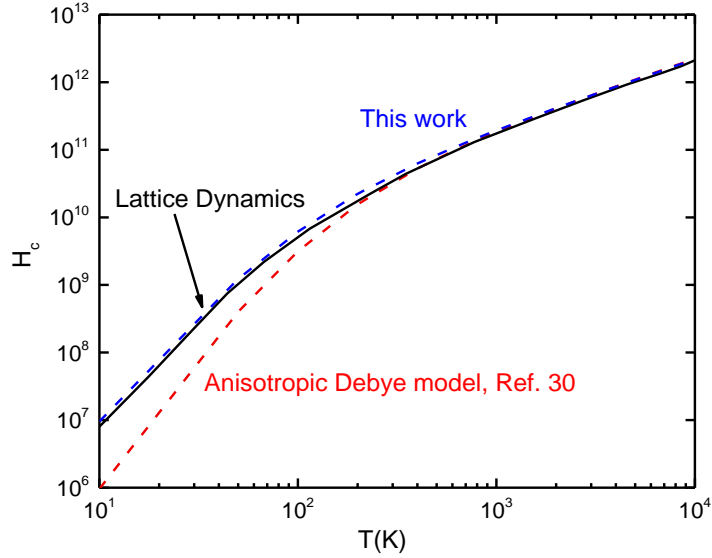

**Figure 3.** Calculated phonon irradiation of graphite (blue dashed line, this work), compared to calculations of a lattice dynamics model (black solid line) and an anisotropic Debye model in Ref. 30 (red dashed line).

Finally, to understand the differences between our model and the anisotropic Debye model by Dames et al.,[30] we compare calculations of the models to measurements of the thermal conductance ($G$) of graphite interfaces[43] and the thermal conductivity ($\Lambda$) of graphite[44]. We note that in the experimental data of $G$, the metal/graphite interfaces are perpendicular to the $c$-axis in graphite. Thus, in the calculations of $G$, we set the $s$-direction in Eqs. (3) and (4) to the $c$-axis.

For the interfacial thermal transport, we calculate $G$ of the Au/graphite and Al/graphite interfaces using the diffuse mismatch model (DMM)[22,31]. Different from previous implementations of DMM, we assume that phonons transmit elastically across the interfaces and allow mode conversion at the interfaces; hence, the transmission probability is a function of phonon frequency ω and not phonon branch $j$. We thus derive the phonon transmission probability from graphite (side 1) to metals (side 2) as

$$\alpha_{1-2}(\omega) = \frac{[h_s(\omega)]_2}{[h_s(\omega)]_1 + [h_s(\omega)]_2} \quad (10)$$

where $h_s(\omega) = \sum_j h_{s,j}(\omega)$ and *s* here refers to the *c*-axis. We insert $\alpha_{1-2}(\omega)$ into Eq. (3) to obtain the DMM calculations, and compare the calculations to experimental data in Fig. 4(a).

Comparing calculations of *G* using our and the Dames' model, we observe a huge discrepancy for the Au/graphite interface and a smaller discrepancy for the Al/graphite interface, for the whole temperature range of 80 K to 400 K, see Fig. 4(a). The discrepancy mainly originates from the inaccurate approximation of the group velocity of long-wavelength ZA phonons in the Dames' model. Due to the elastic phonon transport across most interfaces, heat transport across the metal/graphite interfaces is dominated by long-wavelength phonons, limited by the highest energy of phonons in metals, $\omega_{max}$. According to Eq. (10), three branches have same $\alpha_{1-2}(\omega)$, which is only a function of phonon frequency $\omega$. As a result, for the metal/graphite interfaces, the contribution of each branch to *G* is proportional to $h_{c,j}(\omega)$ of graphite, see Eq. (3). To identify the contributions of each branch, we plot the calculated $h_{c,j}(\omega)$ of the TA, TL1 and TL2 branches of graphite, and label the $\omega_{max}$ of Au and Al (i.e., $\omega_{max,Au}$ and $\omega_{max,Al}$), in Fig. 4(b). As shown in the figure, when $\omega < \omega_{max}$ of Au and Al, $h_{c,j}(\omega)$ of the TL2 branch is much larger than $h_{c,j}(\omega)$ of other branches, suggesting that heat is carried across the metal/graphite interfaces predominantly by ZA phonons and LA phonons along *c*-axis. In this regard, we note that the Dames' model underestimates $h_{c,j}(\omega)$ of the TL2 branch by more than an order of magnitude, see Fig. 4(b), owing to the overestimated "secant" velocity in the Dames' model. (Note that $h_{c,j}(\omega) \propto v_{Z,1}^{-2}$ in the low frequency regime, as summarized in Table 2.) The underestimated $h_{c,j}(\omega)$ of ZA phonons leads to a substantially lower *G* of the metal/graphite interfaces predicted by the Dames' model, see Fig. 4(a).

Fig. 4(b) also suggests that as the phonon energy in metals (i.e., $\omega_{max}$) increases, the difference between calculations of our and the Dames' models decreases. When the phonon energy is sufficiently high, the Dames' model overestimates the contributions of short-wavelength, high-frequency phonons in the TA and TL1 branches, see Fig. 4(b). The overestimation of the contribution of high-frequency phonons compensates for the underestimation of the contribution of low-frequency phonons in the TL2 branch, resulting in a smaller difference in the calculations of *G*, see the calculations for the Al/graphite interface in Fig. 4(a).

Interestingly, differences between calculations of our and the Dames' models are not restricted to the low temperature regime, even though the difference in the calculated $H_c$ (Fig. 3) is observed only at low temperatures. The reason is that for *G* of the metal/graphite interfaces, only long-wavelength phonons with $\omega < \omega_{max}$ of the metal contribute to the interfacial thermal transport owing to the low frequency of phonons in metals, while all phonon modes in graphite contribute to $H_c$. Thus, our model is particularly crucial for studies of heat transport across graphene and graphite interfaces, even at elevated temperatures.

We also note an important consequence of the underestimation of long-wavelength ZA phonons. From Fig. 4(a), comparison of calculations by the Dames' model might erroneously suggest that DMM could well describe heat transport across the Au/graphite interface, but not other graphite interfaces such as the Al/graphite interface. Calculations using our model, however, suggest that heat transport across graphite interfaces is about 50 % of the radiation limit, see more discussion in Ref. 35.

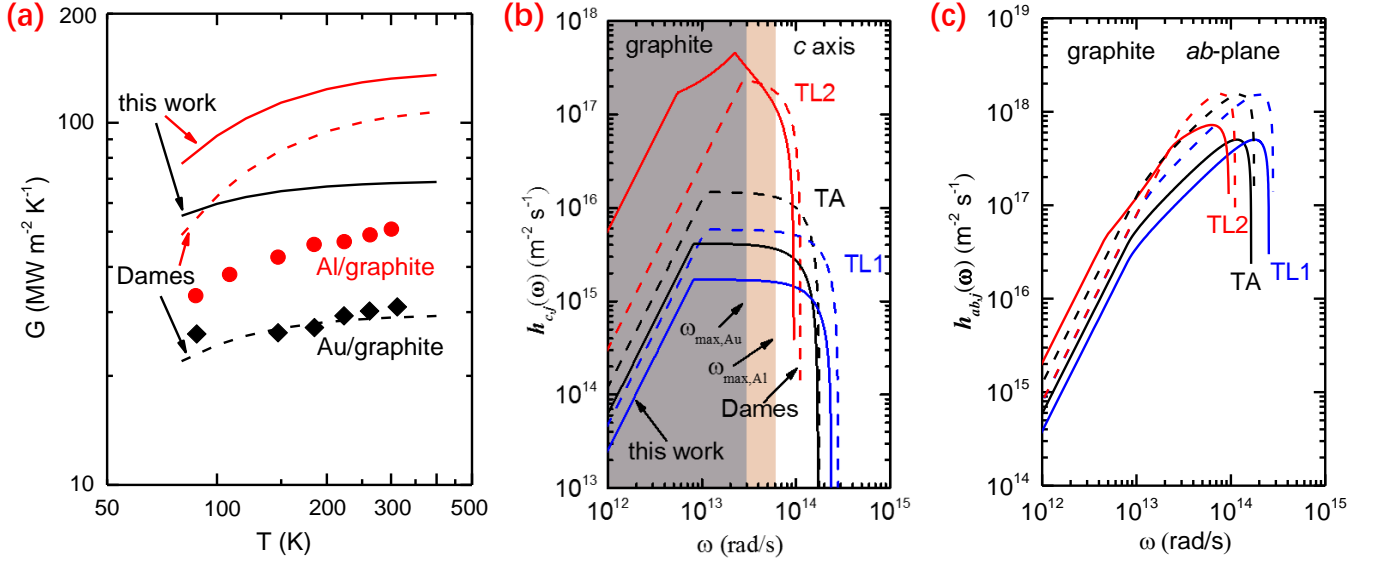

**Figure 4. (a)** Comparison of the DMM calculations using our model (solid lines) and the Dames' model (dashed lines) to the experimental data (solid symbols, Ref. 43) of the thermal conductance $G$ of the Au/graphite (black) and Al/graphite (red) interfaces. **(b)** $h_{c,j}(\omega)$ and **(c)** $h_{ab,j}(\omega)$ of TA (black), TL1 (blue) and TL2 (red) branches for graphite by our model (solid lines) and the Dames' model (dashed lines). The $\omega_{max}$ of Au and Al (i.e., $\omega_{max,Au}$ and $\omega_{max,Al}$) are labelled in (b).

In Fig. 5(a), we also compare calculations of the in-plane ($\Lambda_{ab}$) and through-plane ($\Lambda_c$) thermal conductivity of graphite by our and the Dames' models, to measurements of $\Lambda_{ab}$ and $\Lambda_c$ of graphite in the literature,[44] to illustrate the differences in the calculations of both models. In the calculations, we assume that the average phonon relaxation times of all three branches are given by $\tau_{ab}(\omega) = A_{ab}\omega^{-2}T^{-1}$ for heat transport in the *ab*-plane, and $\tau_c(\omega) = A_c\omega^{-2}T^{-1}$ for heat transport along the *c*-axis direction, respectively, where $A_{ab} = 3.2 \times 10^{19}$ rad$^2$ K s$^{-1}$ and $A_c = 3 \times 10^{17}$ rad$^2$ K s$^{-1}$ are free parameters obtained by fitting calculations of our model to the experimental data, see Fig. 5(a). We then apply the same relaxation times to calculate the anisotropic thermal conductivity of graphite using the Dames' model.

Interestingly, we observe a large discrepancy of more than an order of magnitude for the calculations of $\Lambda_c$, and a much smaller discrepancy for calculations of $\Lambda_{ab}$, see Fig. 5(a). Similar to the calculations of $G$, the large discrepancy in $\Lambda_c$ originates from underestimation of the contribution of long-wavelength ZA phonons in the Dames' model. For the thermal conductivity, when the same $\tau_s(\omega)$ is assumed for all phonon branches, Eq. (1) suggests that the contribution of each branch to $\Lambda_s$ is proportional to $g_{s,j}(\omega)$. Thus, to determine the role of each branch, we plot $g_{c,j}(\omega)$ along the $c$-axis and $g_{ab,j}(\omega)$ in the $ab$-plane for all phonon branches in Fig. 5(b) and 5(c), respectively. We find that while all three branches have comparable contributions to $\Lambda_{ab}$ (Fig. 5(c)), the ZA phonons and the LA phonons along $c$-axis contribute dominantly to $\Lambda_c$, see the large $g_{c,j}(\omega)$ values of the TL2 branch in Fig. 5(b). The observation suggests that ZA phonons and the LA phonons along $c$-axis contribute the most to the heat transport along $c$-axis for layered materials, provided that three branches have similar relaxation times. A similar conclusion has been reached by Ref. 32 based on calculations of the minimum thermal conductivity of WSe$_2$. Similar to calculations of $h_{c,j}(\omega)$, the Dames' model underestimate $g_{c,j}(\omega)$ of the TL2 branch by more than an order of magnitude, see Fig. 5(b), due to the overestimated "secant" velocity of ZA phonons in the Dames' model. This leads to hugely underestimated $\Lambda_c$, see Fig. 5(a).

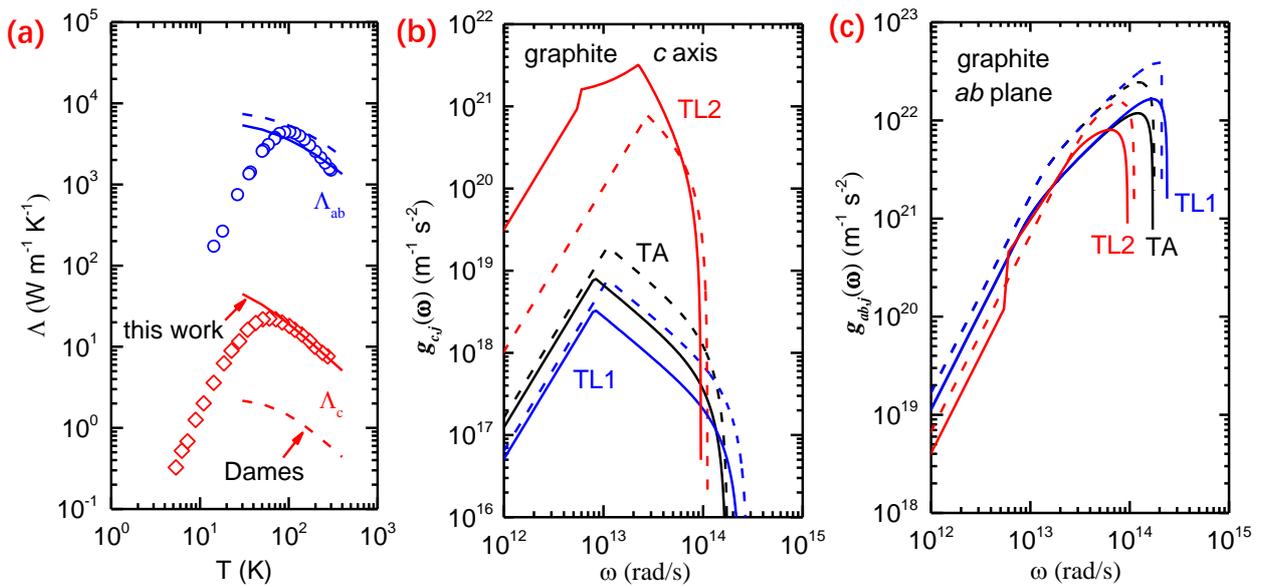

**Figure 5. (a)** Experimental data of anisotropic thermal conductivity along the $c$-axis (red open

diamonds) and in the *ab*-plane (blue open circles), compared to calculations based on our model (solid lines) and the Dames' model (dashed lines). Here, we fit our calculations to the experimental data to derive the fitting parameters $A_{ab}$ and $A_c$, see the main text for details. **(b)** $g_{c,j}(\omega)$ and **(c)** $g_{ab,j}(\omega)$ of TA (black), TL1 (blue) and TL2 (red) branch for graphite from our model (solid lines) and the Dames' model (dashed lines).

For the convenience of other researchers to adopt this approach for calculations of thermal conductivity and thermal conductance, we plot the $h_{ab,j}(\omega)$, $h_{c,j}(\omega)$, $g_{ab,j}(\omega)$ and $g_{c,j}(\omega)$ of graphite in Figs. 3 and 4, and MoS$_2$ in Fig. S1. We notice a few interesting observations. First, in the low frequency regime, we find that all $h_{s,j}(\omega)$ and $g_{s,j}(\omega)$ are proportional to $\omega^2$, as expected. Second, we find that $h_{s,j}(\omega)$ or $g_{s,j}(\omega)$ in one direction (*ab*-plane or *c*-axis) is significantly affected by the group velocities of phonons propagating in the perpendicular direction (*c*-axis or *ab*-plane), an effect known as phonon focusing. The feature can be concluded from the expressions in Tables 1 and 2.

In summary, we develop an anisotropic model with truncated linear dispersions for LA and TA phonons, and a piecewise linear approximation for ZA phonons. In our anisotropic model, we derive the speeds of sounds and the cutoff frequencies directly from the phonon dispersion; thus our assumed phonon dispersion matches the real dispersion well. As a result, our model could accurately calculate the $h_{s,j}(\omega)$ and $g_{s,j}(\omega)$ (i.e., vDOS and v$^2$DOS in other literatures) of layered materials, as exemplified by calculations of phonon irradiation of graphite using our model that agrees excellently well with the molecular dynamic calculations over a wide temperature range from 10 K to 10000 K. We compare the calculations of the thermal conductance or metal/graphite interfaces and the anisotropic thermal conductivity of graphite, by our and the Dames' model, to demonstrate the importance of the piecewise linear approximation of ZA phonons in our model. We find that our model is particularly crucial for calculations of heat transport across the basal planes of layered materials, especially if the heat transport is dominated by low frequency phonons. Our model is simple to implement and thus could be used for quick evaluation of experimental data in the studies of thermal transport in layered materials.

**Acknowledgements**

This work is supported by the Singapore Ministry of Education Academic Research Fund Tier 1 FRC Project FY2016.

**Supplementary Materials for**

# Anisotropic model with truncated linear dispersion for lattice and interfacial thermal transport in layered materials

**Hongkun Li, Weidong Zheng, Yee Kan Koh***

*Department of Mechanical Engineering, National University of Singapore, Singapore 117576*

*Corresponding Author. Email: mpekyk@nus.edu.sg

# 1. Derivation of isoenergy surface equation for TL2 branch

At the low frequency regimes ($\omega < \omega_{Z,1}$), the equation of the isoenergy surface for TL2 branch is similar to that of TA and TL1 branches. The major and minor axes for TL2 at a certain frequency $\omega$ are $k_{ab,\omega} = \omega/v_{Z,1}$ (ZA phonons in $ab$-plane) and $k_{c,\omega} = \omega/v_{L,c}$ (LA phonons along $c$-axis) respectively.[1,2] Thus, the isoenergy surface for low frequency phonons ($\omega < \omega_{Z,1}$) is

$$\frac{v_{Z,2}^2 k_{ab}^2}{\omega^2} + \frac{v_{L,c}^2 k_c^2}{\omega^2} = 1$$

At high frequency regimes ($\omega_{Z,1} < \omega < \omega_{Z,2}$), the minor (or major) axis along $c$ can be easily determined to be same as $k_{c,\omega} = \omega/v_{L,c}$. However, the dispersion in $ab$ plane is composed of two linear functions. According to the dispersion relations of ZA phonons, for a certain frequency of $\omega$, the $k_{ab,\omega}$ can be determined from $k_{Z,1} v_{Z,1} + (k_{ab,\omega} - k_{Z,1}) v_{Z,2} = \omega$. The major axis (or minor) in $ab$ plane is $k_{ab,\omega} = (\omega - \Delta\omega)/v_{Z,2}$, where $\Delta\omega = k_{Z,1}(v_{Z,1} - v_{Z,2})$. Thus, at high frequency regimes ($\omega_{Z,1} < \omega < \omega_{Z,2}$), the isoenergy surface can be written as

$$\frac{v_{Z,2}^2 k_{ab}^2}{(\omega - \Delta\omega)^2} + \frac{v_{L,c}^2 k_c^2}{\omega^2} = 1$$

# 2. Evaluations of $h_{ab,j}(\omega)$, $h_{c,j}(\omega)$, $g_{ab,j}(\omega)$ and $g_{c,j}(\omega)$ for TA, TL1 and TL2 branches

We start the evaluations for TL2 branches, because we can find that TA and TL1 branches are the special case of TL2. After we have the expressions for TL2, then we set $v_{Z,1} = v_{Z,2}$ ($\Delta\omega = 0$). We can have the expressions for TA and TL1 branches.

For the group velocity, we have

$$\mathbf{v}_{g,j} = \left( \frac{\partial \omega}{\partial k_a}, \frac{\partial \omega}{\partial k_b}, \frac{\partial \omega}{\partial k_c} \right), \quad \|\mathbf{v}_{g,j}\| = \sqrt{\left(\frac{\partial \omega}{\partial k_a}\right)^2 + \left(\frac{\partial \omega}{\partial k_b}\right)^2 + \left(\frac{\partial \omega}{\partial k_c}\right)^2}$$

According to Eqs. (2) and (4) in the main text, given that $g_{s,j}(\omega)$ and $h_{s,j}(\omega)$ ($j$ denotes TL2 in this case) both have the parts of surface integral of $\iint \frac{dS_\omega}{\|\mathbf{v}_{g,j}\|}$, we can simply this part as

$$\iint \frac{dS_\omega}{\|\mathbf{v}_{g,j}\|} = \iint \frac{\sqrt{1+\left(\frac{\partial k_c}{\partial k_a}\right)^2 + \left(\frac{\partial k_c}{\partial k_b}\right)^2}}{\sqrt{\left(\frac{\partial \omega}{\partial k_a}\right)^2 + \left(\frac{\partial \omega}{\partial k_b}\right)^2 + \left(\frac{\partial \omega}{\partial k_c}\right)^2}} dk_a dk_b = \iint \frac{\sqrt{1+\left(\frac{\partial k_c}{\partial \omega}\frac{\partial \omega}{\partial k_a}\right)^2 + \left(\frac{\partial k_c}{\partial \omega}\frac{\partial \omega}{\partial k_b}\right)^2}}{\sqrt{\left(\frac{\partial \omega}{\partial k_a}\right)^2 + \left(\frac{\partial \omega}{\partial k_b}\right)^2 + \left(\frac{\partial \omega}{\partial k_c}\right)^2}} dk_a dk_b$$

$$= \iint \frac{\sqrt{\left(\frac{\partial k_c}{\partial \omega}\right)^2 \left(\left(\frac{\partial \omega}{\partial k_c}\right)^2 + \left(\frac{\partial \omega}{\partial k_a}\right)^2 + \left(\frac{\partial \omega}{\partial k_b}\right)^2\right)}}{\sqrt{\left(\frac{\partial \omega}{\partial k_a}\right)^2 + \left(\frac{\partial \omega}{\partial k_b}\right)^2 + \left(\frac{\partial \omega}{\partial k_c}\right)^2}} dk_a dk_b = \iint \frac{\partial k_c}{\partial \omega} dk_a dk_b$$

**Thus, for TL2 branches,** $\omega < \omega_{Z,1}$, the isoenergy surface and effective FBZ boundary are

$$\frac{v_{Z,2}^2 k_{ab}^2}{\omega^2} + \frac{v_{L,c}^2 k_c^2}{\omega^2} = 1$$

$$\frac{k_{ab}^2}{k_{Z,2}^2} + \frac{k_c^2}{k_{L,c}^2} = 1$$

We use $g_{ab,j}(\omega)$ as an example to show the derivations. Along $c$-axis and in $ab$-plane, the unit vector $\hat{\mathbf{s}}$ are $(0, 0, 1)$ and $(1, 0, 0)$ respectively. Also we apply the polar coordinate substitution ($k_a = \rho \cos\varphi$, $k_b = \rho \sin\varphi$) to simplify the evaluation. Please note that for $g_{s,j}(\omega)$, there should be a factor of 2 accounting for the integral for upper and lower hemisphere of FBZ, while for $h_{s,j}(\omega)$, there is no need due to the integral for $h_{s,j}(\omega)$ is over half of the incident FBZ.

$$g_{ab,j}(\omega) = \frac{1}{8\pi^3} \iint \frac{\left(\mathbf{v}_{g,j} \cdot \hat{\mathbf{s}}\right)^2}{\|\mathbf{v}_{g,j}\|} dS_\omega = \frac{1}{8\pi^3} \iint \left(\frac{\partial \omega}{\partial k_a}\right)^2 \frac{\partial k_c}{\partial \omega} dk_a dk_b = \frac{1}{4\pi^3} \iint \frac{\rho^3 \cos^2\varphi \, v_{Z,1}^4}{\omega v_{L,c} \sqrt{(\omega^2 - \rho^2 v_{Z,1}^2)}} d\rho d\varphi$$

In which the domain of $\varphi$ and $\rho$ are $0 \leq \varphi \leq 2\pi$ and $0 \leq \rho \leq \omega/v_{Z,1}$, thus,

$$g_{ab,j}(\omega) = \frac{1}{4\pi^3} \iint \frac{\rho^3 \cos^2\varphi \, v_{Z,1}^4}{\omega v_{L,c} \sqrt{(\omega^2 - \rho^2 v_{Z,1}^2)}} d\rho d\varphi = \frac{1}{4\pi^2} \int \frac{\rho^3 v_{Z,1}^4}{\omega v_{L,c} \sqrt{(\omega^2 - \rho^2 v_{Z,1}^2)}} d\rho$$

$$= \frac{1}{4\pi^2} \frac{v_{Z,1}^4}{\omega v_{L,c}} \left[ \frac{(\omega^2 - \rho^2 v_{Z,1}^2)^{\frac{3}{2}}}{3 v_{Z,1}^4} - \frac{\omega^2 (\omega^2 - \rho^2 v_{Z,1}^2)^{\frac{1}{2}}}{v_{Z,1}^4} \right]\Bigg|_0^{\omega/v_{Z,1}} = \frac{1}{6\pi^2 v_{L,c}} \omega^2$$

**For TL2 branches, $\omega_{Z,1} < \omega < \omega_{L,c}$**, the isoenergy surface and effective FBZ boundary are

$$\frac{v_{Z,2}^2 k_{ab}^2}{(\omega - \Delta\omega)^2} + \frac{v_{L,c}^2 k_c^2}{\omega^2} = 1$$

$$\frac{k_{ab}^2}{k_{Z,2}^2} + \frac{k_c^2}{k_{L,c}^2} = 1$$

The $g_{ab,j}(\omega)$ can be expressed as

$$g_{ab,j}(\omega) = \frac{1}{8\pi^3} \iint \frac{(\mathbf{v}_{g,j} \cdot \hat{\mathbf{s}})^2}{\|\mathbf{v}_{g,j}\|} dS_\omega = \frac{1}{8\pi^3} \iint \left(\frac{\partial \omega}{\partial k_a}\right)^2 \frac{\partial k_c}{\partial \omega} dk_a dk_b$$

$$= \frac{1}{4\pi^3} \iint \frac{\rho^3 \cos^2\varphi \, v_{Z,2}^4 \omega^2}{\rho^2 v_{Z,2}^2 \Delta\omega + (\omega - \Delta\omega)^3} \frac{1}{v_{L,c} \sqrt{(\omega - \Delta\omega)^2 - \rho^2 v_{Z,2}^2}} d\rho d\varphi$$

In which the domain of $\varphi$ and $\rho$ are $0 \leq \varphi \leq 2\pi$ and $0 \leq \rho \leq (\omega - \Delta\omega)/v_{Z,2}$, but in this case, it becomes difficult to integrate over $\rho$ to find a analytical solution. So we only integrate over $\varphi$ and do the integration of $\rho$ using numerical approach.

$$g_{ab,j}(\omega) = \frac{1}{4\pi^3} \iint \frac{\rho^3 \cos^2\varphi \, v_{Z,2}^4 \omega^2}{\rho^2 v_{Z,2}^2 \Delta\omega + (\omega - \Delta\omega)^3} \frac{1}{v_{L,c} \sqrt{(\omega - \Delta\omega)^2 - \rho^2 v_{Z,2}^2}} d\rho d\varphi$$

$$= \frac{1}{4\pi^2} \int \frac{\rho^3 v_{Z,2}^4 \omega^2}{\rho^2 v_{Z,2}^2 \Delta\omega + (\omega - \Delta\omega)^3} \frac{1}{v_{L,c} \sqrt{(\omega - \Delta\omega)^2 - \rho^2 v_{Z,2}^2}} d\rho$$

**For TL2 branches, $\omega_{L,c} < \omega < \omega_{Z,2}$**, the isoenergy surface and effective FBZ boundary are same as above

$$\frac{v_{Z,2}^2 k_{ab}^2}{(\omega-\Delta\omega)^2} + \frac{v_{L,c}^2 k_c^2}{\omega^2} = 1$$

$$\frac{k_{ab}^2}{k_{Z,2}^2} + \frac{k_c^2}{k_{L,c}^2} = 1$$

Thus, the evaluation for the $g_{ab,j}(\omega)$ is same as second case, that is

$$g_{ab,j}(\omega) = \frac{1}{4\pi^2} \int \frac{\rho^3 v_{Z,2}^4 \omega^2}{\rho^2 v_{Z,2}^2 \Delta\omega + (\omega-\Delta\omega)^3} \frac{1}{v_{L,c}\sqrt{(\omega-\Delta\omega)^2 - \rho^2 v_{Z,2}^2}} d\rho$$

The only difference is that, in this case, parts of the isoenergy surface lie outside of the effective FBZ, and thus should be excluded from the integration. The domain of $\rho$ is $\rho_{min} \leq \rho \leq \rho_{max}$, where $\rho_{min} = k_{Z,2}\sqrt{\frac{\omega^2(\omega-\Delta\omega)^2 - \omega_{L,c}^2(\omega-\Delta\omega)^2}{k_{Z,2}^2 v_{Z,2}^2 \omega^2 - \omega_{L,c}^2(\omega-\Delta\omega)^2}}$, $\rho_{max} = (\omega - \Delta\omega)/v_{Z,2}$, while the domain of $\varphi$ is still $0 \leq \varphi \leq 2\pi$.

## 3. Plots of $h_{ab,j}(\omega)$, $h_{c,j}(\omega)$, $g_{ab,j}(\omega)$ and $g_{c,j}(\omega)$ for MoS$_2$

For the convenience of other researchers to adopt this approach for calculations of thermal conductivity and thermal conductance, we also plot the $h_{ab,j}(\omega)$, $h_{c,j}(\omega)$, $g_{ab,j}(\omega)$ and $g_{c,j}(\omega)$ of MoS$_2$ in Fig. S1. The input parameters for MoS$_2$ are determined from published phonon dispersion and have been list in Table 3 in main text.

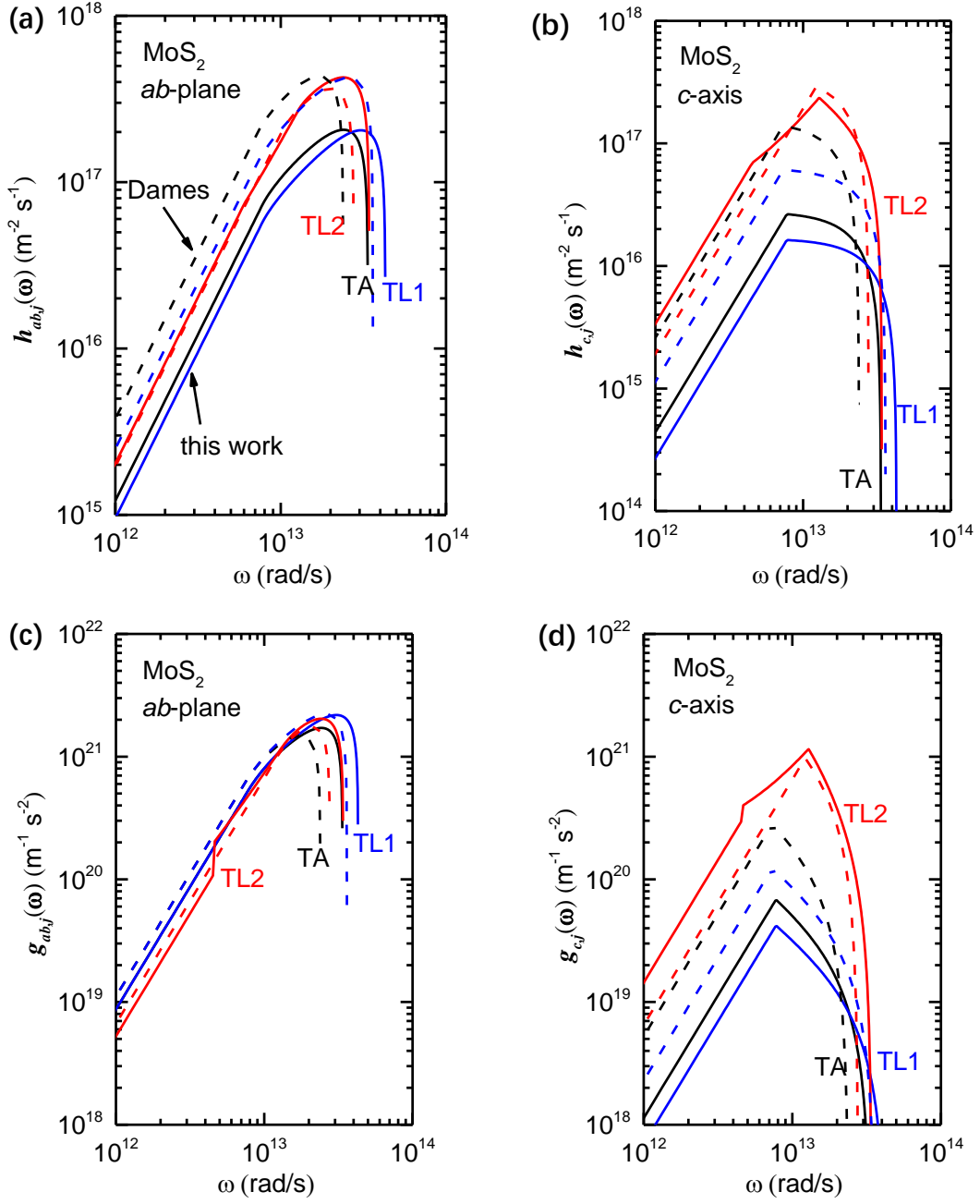

Figure. S1. (a) $h_{ab,j}(\omega)$, (b) $h_{c,j}(\omega)$, (c) $g_{ab,j}(\omega)$ and (d) $g_{c,j}(\omega)$ of TA (black), TL1 (blue) and TL2 (red) branch for MoS$_2$ from our model (solid lines) and Dames' model (dashed lines).

# References

1. Bertram Alexander Auld, Acoustic fields and waves in solids. (Рипол Классик, 1973).
2. Z. Chen, Z. Wei, Y. Chen, and C. Dames,   Physical Review B **87** (12) (2013).